\documentclass[prl,twocolumn,superscriptaddress,showpacs,psfig,showkeys,nofootinbib]{revtex4}
\usepackage{amsmath}
\usepackage{amssymb}
\usepackage{amsfonts}
\usepackage{graphicx}
\usepackage{dcolumn}
\usepackage{epstopdf}
\usepackage{color}
\usepackage[utf8]{inputenc}
\usepackage{amsthm}
\usepackage{fontenc}
\usepackage{bm}
\RequirePackage{slashed}\usepackage{cancel}
\usepackage[normalem]{ulem}

\usepackage{array,booktabs}
\newcolumntype{C}{>{$\displaystyle}c<{$}}

\begin{document}


\title{Catching a glimpse 
of the parton structure of the bound proton
}

\author{Sara Fucini}
\affiliation{ Dipartimento di Fisica e Geologia,
Universit\`a degli Studi di Perugia and Istituto Nazionale di Fisica Nucleare,
Sezione di Perugia, via A. Pascoli, I - 06123 Perugia, Italy}
\author{Sergio Scopetta}
\affiliation{ Dipartimento di Fisica e Geologia,
Universit\`a degli Studi di Perugia and Istituto Nazionale di Fisica Nucleare,
Sezione di Perugia, via A. Pascoli, I - 06123 Perugia, Italy}
\author{Michele Viviani}
\affiliation{
INFN-Pisa, 56127 Pisa, Italy} 
\date{\today}

\begin{abstract}
\vspace{0.5cm}
A new generation of experiments is expected to shed
light on the elusive parton structure of the bound proton.
One of the most promising directions
is incoherent deeply virtual Compton scattering,
which can provide a tomographic view
of the bound proton.
The first measurement 
has been recently performed,
using $^4$He targets at Jefferson Lab. 
In the work presented here,
a rigorous Impulse Approximation
analysis
of this process is proposed.
As ingredients,
state-of-the-art models of the
nuclear spectral function and of the 
parton structure of the struck proton,
together with novel
scattering amplitudes
expressions
for a bound moving nucleon, have been used. 
The overall agreement obtained with the data, good in particular
at high values of the photon virtuality, demonstrates the solidity of the
framework, which is also suitable for further improvements. It is found that
possible big differences between results for the bound proton and 
those for the free one 
could be related to kinematical nuclear effects and not to
modifications of the parton structure.
The analysis demonstrates that the comparison of the results of this approach,
based on a conventional description,
with future precise data, has the potential
to expose exotic quark and gluon effects in nuclei.

\end{abstract}

\maketitle

Several decades ago, the discovery of the European Muon 
collaboration (EMC) effect
in inclusive deep inelastic scattering
off nuclear targets
\cite{Aubert:1983xm}
has shown that
the parton structure of bound nucleons
is modified by the nuclear medium
(see Ref. \cite{Hen:2013oha} for a recent report).
A new generation of planned measurements at
high energy and high luminosity facilities
could provide in the next years, for the
first time, a fully quantitative explanation of the EMC effect
(for a recent report, see, e.g., Ref. 
\cite{Dupre:2015jha,Cloet:2019mql}).
This programme includes the
challenging realization of
semi-inclusive and exclusive experiments
and their complicated 
theoretical description.
Among the most promising directions, nuclear deeply virtual Compton scattering (DVCS), the hard  exclusive leptoproduction of a real photon on a nuclear target, plays a special role. 
In DVCS, the parton structure is encoded in the so called
generalized parton distributions (GPDs)
\cite{gpds}, non perturbative 
quantities providing a wealth of novel information
(for exhaustive reports see, e.g.,
Ref. \cite{Diehl:2003ny}).
Nuclear DVCS could unveil
the presence of non-nucleonic degrees of freedom
\cite{Berger:2001zb}, or
may allow to better understand the distribution of nuclear forces in nuclei
\cite{Polyakov:2018zvc}. Nonetheless, the subject of this letter
is mostly related to the tomography of the
bound proton, i.e., the distribution
of partons with a given longitudinal momentum
in the transverse plane. This is certainly
one of the most exciting information accessible in DVCS through the GPDs formalism
\cite{Burkardt:2000za}.
In nuclei, DVCS can occur
through two different mechanisms, i.e., the coherent one $A(e,e'\gamma)A$, where the target $A$ remains intact recoiling as a whole, and the incoherent one $A(e,e'\gamma p)X$, where the nucleus breaks up and the struck proton is detected, so that its
tomography could be ultimately obtained. The comparison between this information and that obtained for the free proton could provide a pictorial view of the realization of the EMC effect. 
From an experimental point of view, the study of nuclear DVCS requires the very difficult coincidence detection 
of fast photons and electrons
together with slow, intact recoiling protons
or nuclei. 
For this reason, in the first measurement of nuclear DVCS at
HERMES \cite{Airapetian:2009cga}, a clear separation 
was not achieved between the two different DVCS channels.
Nevertheless, recently, for the first time, such a 
separation has been performed by the EG6 experiment
of the CLAS collaboration \cite{eric}, with the 6 GeV electron beam at Jefferson Lab (JLab).
The first data for coherent 
and incoherent DVCS off $^4$He have been published in Refs. \cite{Hattawy:2017woc}
and
\cite{Hattawy:2018liu}, respectively. 
Among few nucleon systems, 
for which a realistic evaluation of 
conventional nuclear effects is possible
in principle,
$^4$He is deeply bound and
represents the prototype of a typical finite nucleus. 
Realistic approaches allow to distinguish conventional nuclear effects
from exotic ones, which could be responsible of the observed EMC behaviour. Without realistic benchmark calculations,  
the interpretation of the data will be hardly conclusive.  In fact, in  Refs. \cite{Hattawy:2017woc,Hattawy:2018liu}, the
importance of new calculations has been addressed, for a
successful interpretation of the collected data and of those planned at JLab
in the next years \cite{Armstrong:2017wfw}. In facts
available estimates, proposed long time ago, correspond
in some cases to different kinematical regions \cite{Guzey:2003jh,Liuti:2005gi}.
We have therefore recently 
performed a successful impulse approximation (IA) analysis of 
coherent DVCS off
$^4$He
\cite{Fucini:2018gso}, obtaining an overall good agreement with the data 
\cite{Hattawy:2017woc}.
In this letter,
we propose 
an analogous analysis for the incoherent
channel, to see to what extent
a conventional description can describe
the recent data \cite{Hattawy:2018liu}
which have the tomography
of the bound proton as the ultimate goal.
One should notice that, in Ref.\cite{Fucini:2018gso}, 
the calculation of the coherent
channel required, as the only theoretical tool, the nuclear GPD.
In the present investigation completely new issues arise, such as
the calculation of the appropriate differential cross sections for
a moving proton in the medium, and the modelling of a diagonal nuclear
spectral function.

\begin{figure}[t]
\includegraphics[scale=0.40,angle=0]{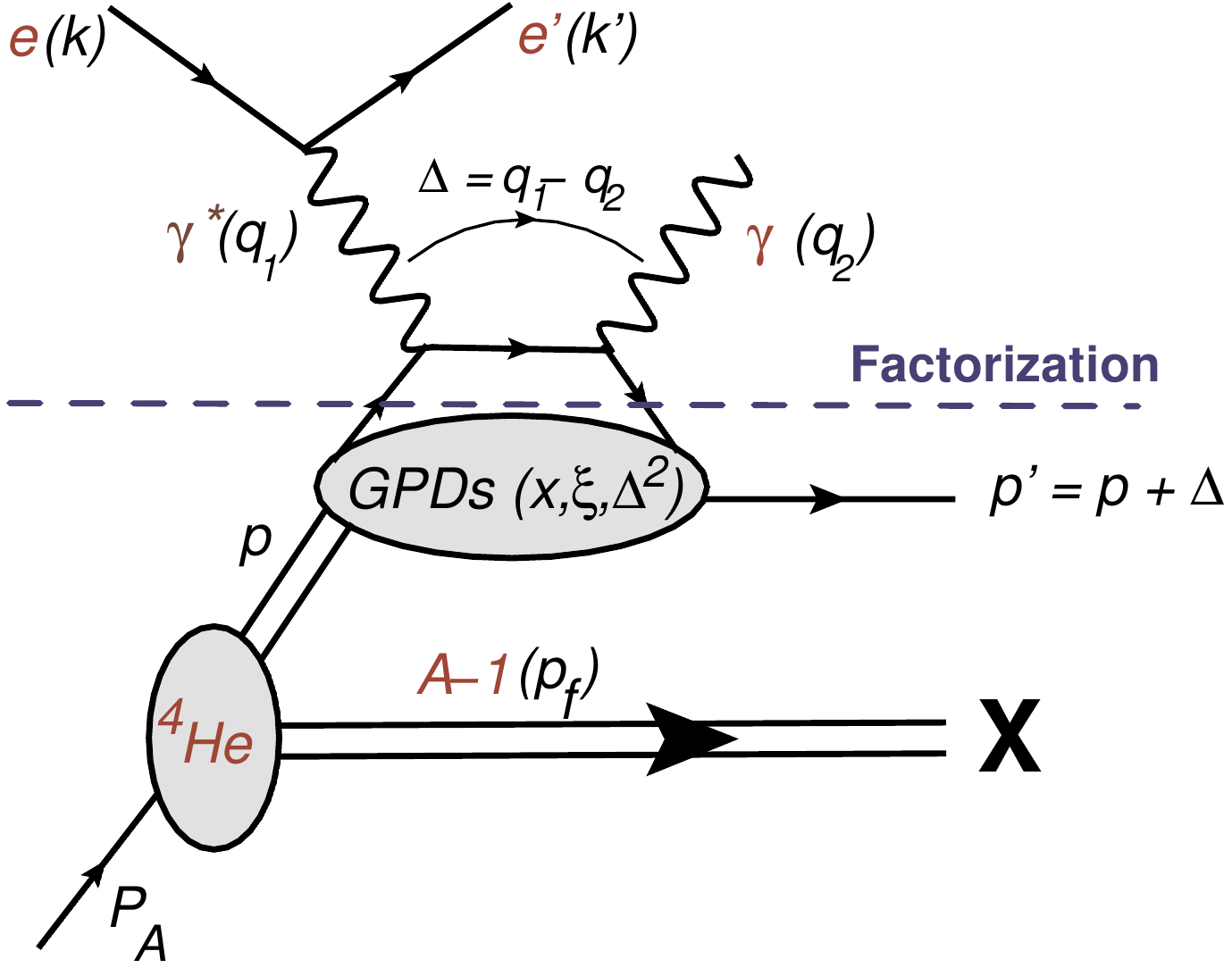}
\caption{(color online)
Incoherent DVCS process off $^4$He in the IA to the handbag approximation.
}
\label{dvcsinco}
\end{figure}

We studied therefore the IA
to the handbag approximation to 
the incoherent DVCS process, $A(e,e' \gamma p')X$, shown in Fig. \ref{dvcsinco}.
It means that we assumed that the process goes through one
quark in one nucleon in $^4$He, i.e., non-nucleonic degrees
of freedom are not considered, and further possible rescattering of the struck detected proton 
with the remnant $X$ is disregarded.
For high enough values of the initial photon virtuality, $Q^2 = -q_1^2 = -(k-k')^2$, IA
usually provides the bulk of nuclear effects in a hard electron scattering process
(see, e.g., Ref. \cite{Slifer:2008re} for an experimental study of the onset of the validity of IA calculations).
Similar expectations hold in the present study, although
only the comparison with data can demonstrate
the validity of the chosen framework.
If $Q^2$
is much larger than $-t = -\Delta^2 = -(p-p')^2 $, the momentum
transferred to the hadronic system
with initial (final) 4-momentum $p(p')$,
the hard vertex of the ``handbag'' 
diagram depicted in Fig. \ref{dvcsinco} 
can be studied perturbatively. 
The soft part is parametrized in terms of GPDs
of the struck proton,
which depend on $\Delta^2$, on
the so-called skewness  $\xi =-{\Delta^+}/{P}^+$,
i.e., the difference in plus momentum fraction between the initial and the final states, and  on $x$,
the average plus momentum fraction of the struck parton 
with respect to the total momentum, not experimentally accessible
(the notation $a^\pm = (a_0 \pm a_3)/\sqrt{2}$ is used). The average four momenta, for photons and protons, are $q=(q_1 +q_2)/2$ and $P = p+p'$, respectively. 
In IA, one also has
$ - \Delta^2 = -(q_1-q_2)^2$, that is,
the momentum transferred to the system coincides with that
transferred to the struck proton.
The reference frame proposed in Ref. \cite{Belitsky:2001ns}, with the target at rest,
the virtual photon with energy $\nu$ moving opposite to the 
$\hat{z}$ axis and the leptonic and hadronic planes of the reaction defining the angle $\phi$,
has been adopted. 
Besides, using energy-momentum conservation, one gets for
the azimuthal angle of the detected proton
the relation $\phi_{p'} = \phi + \phi_e$ and, since in the chosen frame $\phi_e=0$, $\phi_{p'}$ coincides with $\phi$.
A pure DVCS process always 
interferes with
the electromagnetic
Bethe-Heitler (BH) process, 
which produces the same final state $(e' \gamma p')$.
The IA description of the BH process is shown in Fig. 2.
We note in passing that 
the possibility that the real photon is emitted by the initial
nucleus, or by the final X system, has been neglected,
being the BH cross-section approximately proportional to the 
inverse squared mass of the emitter.
With respect to the emission from the electrons, this contribution should
be therefore negligibly small. For this reason,
the experimental collaboration EG6 has not considered this occurrence
in its analysis.
From a theoretical point of view, neglecting these
contributions, gauge invariance is not reproduced. 
Nonetheless we have to point out that
in the present Impulse Approximation analysis gauge invariance is in any 
case not fulfilled and it could be restored at the nuclear level 
only implementing many-body currents. 
These corrections have not been included yet 
in the calculation and they could be more relevant than photon emission 
from nuclear systems.
Since GPDs are not directly measurable,
the experimental way to access their physical content 
exploits the BH-DVCS interference.
In facts,
in the squared amplitude of the process under scrutiny,
\begin{equation}\label{amp}
    \mathcal{A}^2 = T_{DVCS}^2+T_{BH}^2 + \mathcal{I}\,
\end{equation}
in the kinematical region of the performed experiment, the BH mechanism
  is dominating on the DVCS one. 
  By measuring 
  the beam-spin asymmetry (BSA) of the process off an unpolarized (U) target
\begin{equation}\label{ALU}
    A_{LU}= \frac{d\sigma^+ - d \sigma^-}{d \sigma^+ + d \sigma^-}\,,
\end{equation}
where $\pm$ refers to positive/negative longitudinal (L) beam helicity, in a leading-twist analysis
it is possible to
isolate
the BH-DVCS interference
$ \mathcal{I} = 2 \Re e(T_{DVCS}T_{BH}^*)$.
This term is sensitive to the target partonic content, parametrized through GPDs hidden in the so called Compton Form Factors (CFF), appearing
in the $T_{DVCS}$ amplitude. 
We studied therefore the BSA, the observable recently measured at JLab and a workable expression for it
is needed. 

Let us describe our IA calculation of the BSA.
To evaluate Eq. \eqref{ALU}, the cross-section for a DVCS process occurring off a bound moving proton in $^4$He is required.
\begin{figure}[t]
\hspace{-.5cm}
\includegraphics[scale=0.3,angle=0]{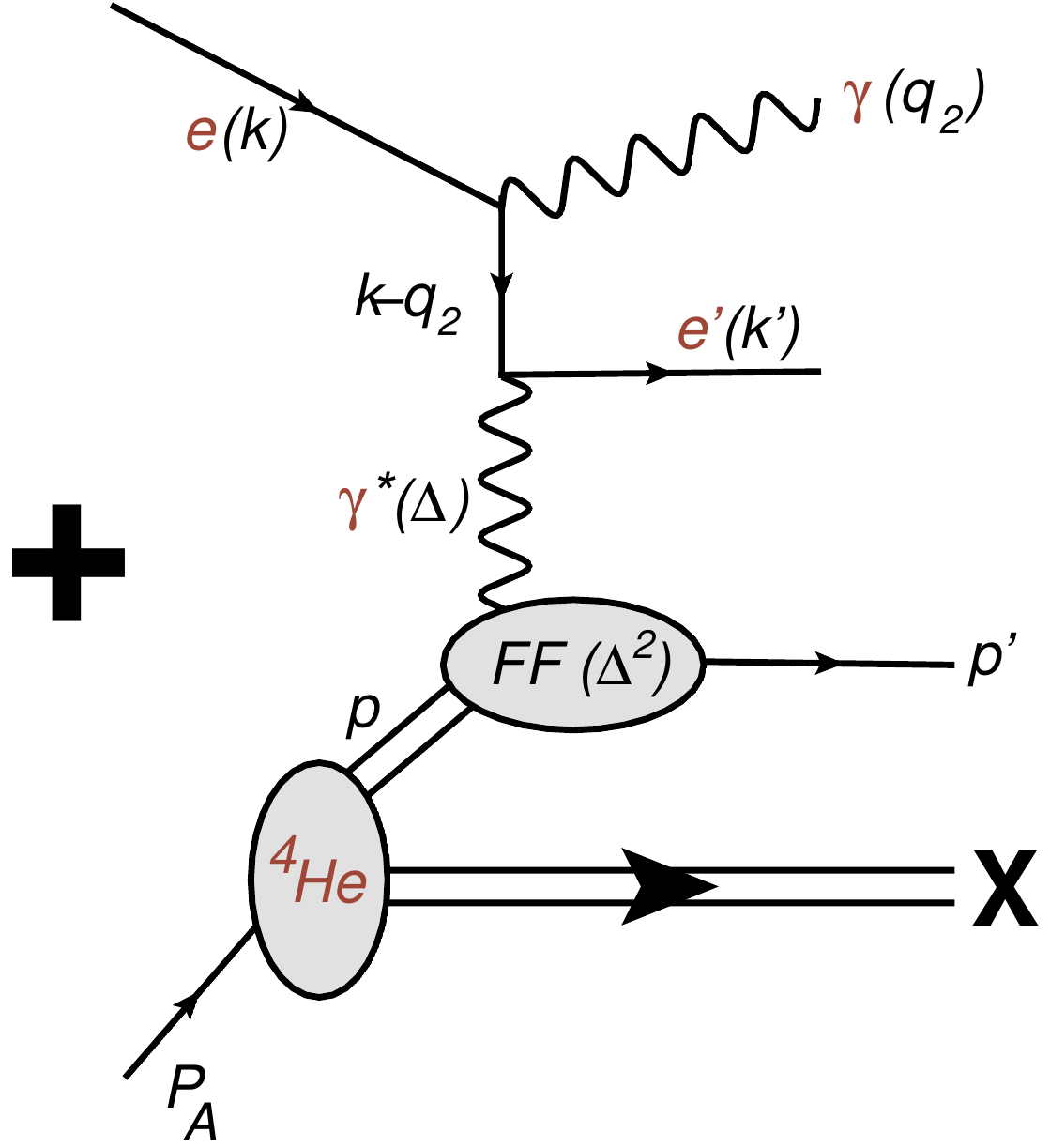}
\hspace{0.6cm}
\includegraphics[scale=0.3,angle=0]{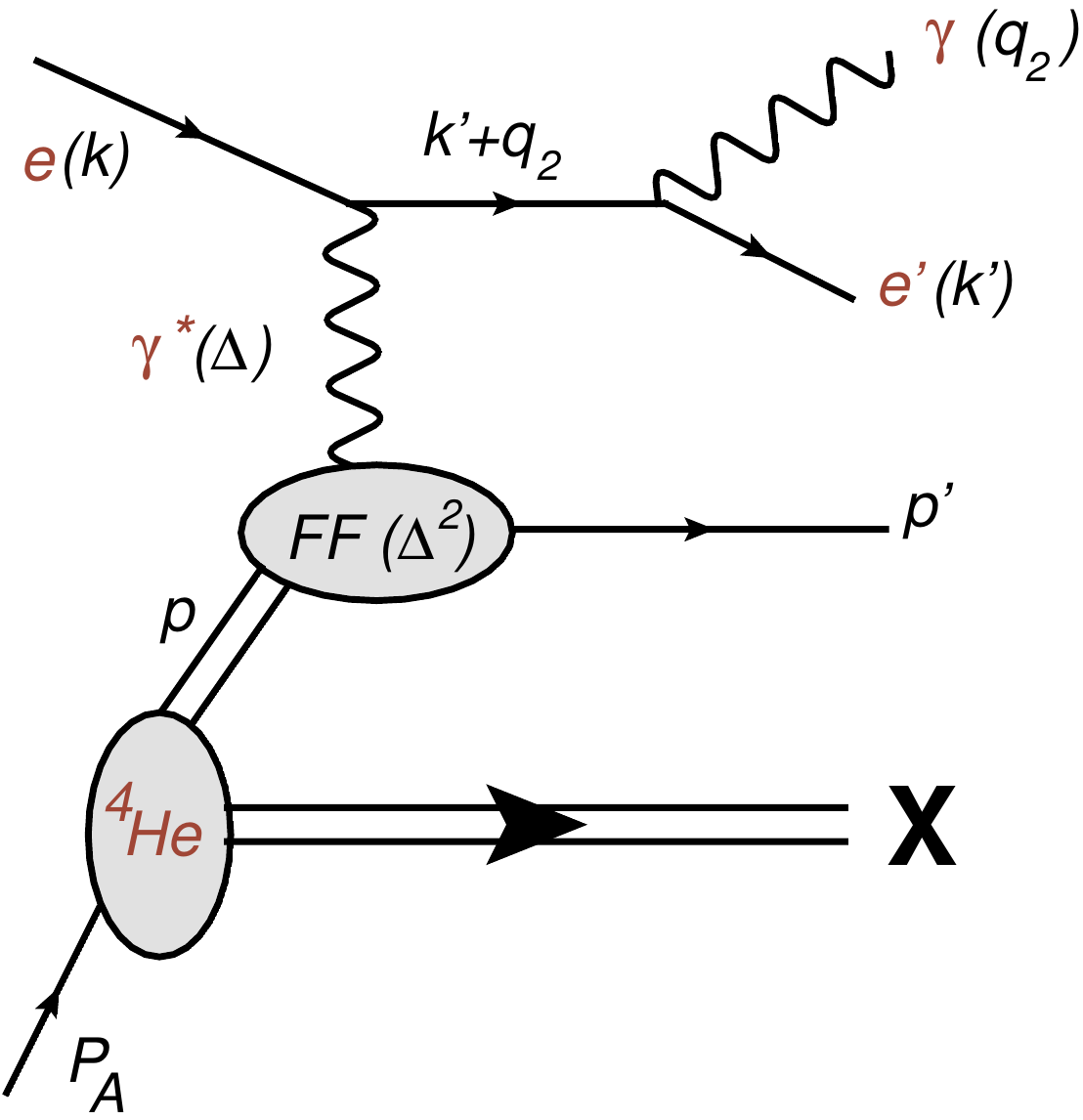}
\caption{(color online) The Bethe Heitler process in IA.}
\label{dvcshb}
\end{figure}
In our IA approach,
the off-shellness of the initial bound proton is purely kinematical, i.e., the energy of the struck proton
is obtained from energy conservation and reads
\begin{equation}
    p_0 = M_A  - \sqrt{M_{A-1}^{*2} + \vec{p}^2}\simeq M - E - T_{rec}\,, 
\end{equation}
where we define the removal energy $E = M^*_{A-1} + M - M_A = \epsilon^*_{A-1}+ |E_A| - |E_{A-1}|$  in terms of the binding energy (mass) of $^4$He and of the 3-body system, $E_A$ ($M_A$) and $E_{A-1}$
($M_{A-1}^*$), respectively, and of the excitation energy
of the recoiling system, $\epsilon^*_{A-1}$. Finally,  $T_{rec}$ is the kinetic energy of the recoiling  $3-$body system and $M$ is the proton mass.
In this way, after a straightforward but lengthy analysis,
which will be shown elsewhere \cite{tobe}, one finds a
complicated convolution formula for the cross section, which can be sketched as follows
\begin{equation}\label{cross}
    d\sigma_{Inc}^\pm= 
    \int_{exp} dE \, d{\vec p} \,
    \frac{p\cdot k}{p_0 \, |\vec k|}
 P^{^4He}(\vec{p},E)\, d\sigma^{\pm}_b(\vec p, E,K)\,,
\end{equation}
in terms of the nuclear spectral function $ P^{^4He}(\vec{p},E) $ and of the cross section for a DVCS process off a bound proton,
$d\sigma^{\pm}_b$. The integral on the removal energy refers to both discrete and continuous energy spectra of $^4$He.
In Eq. \eqref{cross}, $K$ is the set of kinematical variables $\{x_B=Q^2/(2 M \nu),Q^2,t,\phi\}$.
The range of $K$ accessed
in the experiment fixes the proper
energy and momentum
integration space, denoted
as $exp$.
From Eq. \eqref{cross} we get
the measured differential cross section, 
appearing in Eq. \eqref{ALU},
\begin{align}
d \sigma^\pm \equiv  \frac{d\sigma^\pm_{Inc} }{d{x}_BdQ^2 d{\Delta}^2 d\phi}&= \int_{exp} dE \, d{\vec p} \,P^{^4He}(\vec {p},E)
\\
  & \times |\mathcal{A}^{\pm}({\vec p}, E ,K)|^2
g(\vec{p},E,K) \nonumber \,,
\end{align}
where  $g(\vec p,E,K)$ is a complicated function
which arises from the integration over the phase space and includes also the flux factor ${p\cdot k}/({p_0 \,|\vec  k |})$ in Eq. \eqref{cross}. 
This latter term comes from the fact that one has at disposal only  non-relativistic wave functions to evaluate the spectral function. This implies also that either the number or the momentum sum rule is slightly
violated. Such a  problem could be solved with a Light Front approach, as proposed in Ref. \cite{DelDotto:2016vkh} for a $3-$body system.
The BSA assumes the schematic form
\begin{equation}\label{alu_ratio}
A_{LU}^{Incoh} (K) = 
 \frac
 { \mathcal{I}^{^4He} (K) } 
 { T_{BH}^{2\, \, ^4He} (K) 
 }\,,
\end{equation}
where
\begin{equation}\label{nume}
\mathcal{I}^{^4He} (K) =
\int_{exp} dE \, d \vec p \, 
{ {P^{^4He}(\vec p, E )}} 
\, g(\vec p,E,K)\, 
{{\mathcal{I}(\vec p, E, K)}} \,,
\nonumber
\end{equation}
\begin{eqnarray}\label{deno}
T_{BH}^{2\,\,^4He} (K) & = &
\int_{exp} dE \, d \vec p \, {{P^{^4He} (\vec p, E )}} 
\, g(\vec p, E,K)\,\nonumber \\ 
& \times &
{{T_{BH}^2 (\vec p, E, K)}}\,,
\end{eqnarray}
where $\mathcal{I}$ and
$T_{BH}^2$ refer to a moving bound nucleon and
generalize the Fourier decomposition of the DVCS cross section off a proton at rest, at leading twist, derived in
Ref. \cite{Belitsky:2001ns}.
The new expressions for $\mathcal{I}$ and
$T_{BH}^2$ will be presented elsewhere \cite{tobe}; here,
without going into technical details, we summarize 
the structure of the different contributions.
For the BH part, we considered the full sum of azimuthal harmonics, i.e
$T_{BH}^2 = c_0^{bound} + c_1^{bound}\cos \phi + c_2^{bound} \cos(2 \phi )$,
where the coefficients $c_i^{bound}$ contain the Dirac and Pauli form factors (FFs).
This decomposition
is driven
by the explicit form of the BH propagators
shown in Fig. \ref{dvcshb}.
We stress that in the present IA approach no nuclear modifactions occur for the FFs of the bound proton.
As for the interference term, we considered the leading twist contribution, so that terms explicitly proportional to $\Delta^2/Q^2$ \cite{htwist}
have been neglected while corrections proportional to  $\epsilon^2$, with  $\epsilon=  2 M x_B/{Q}$, accounting for target mass corrections, have been considered.  
In the numerator of Eq. \eqref{ALU} only the
term accounting for the beam polarization
is selected, where
the dependence on the parton structure of the bound proton is hidden in the imaginary part of the CFF  $\mathcal{H}$.
In the kinematics of interest, this quantity 
can be 
expressed in terms of only one GPD 
of the bound proton, $H(x,\xi,\Delta^2)$,  
according to
$\Im m \, \mathcal{H}(\xi',t) = H(\xi',\xi',t)- H(-\xi', \xi', t)$. We notice that
the off-shellness affects the proton parton structure, since the GPDs have to be evaluated for a skewness $ \xi' = {Q^2}/({2 P \cdot q}) $ given in terms of the 4-momenta of the proton and the photons. The modification at partonic level is due to this rescaling of the skewness that, for a proton at rest, reduces to $ \xi \simeq x_B/(2 - x_B) $.
\begin{figure*}[t]
\hspace{-.5cm}
\includegraphics[scale=0.85,angle=0]{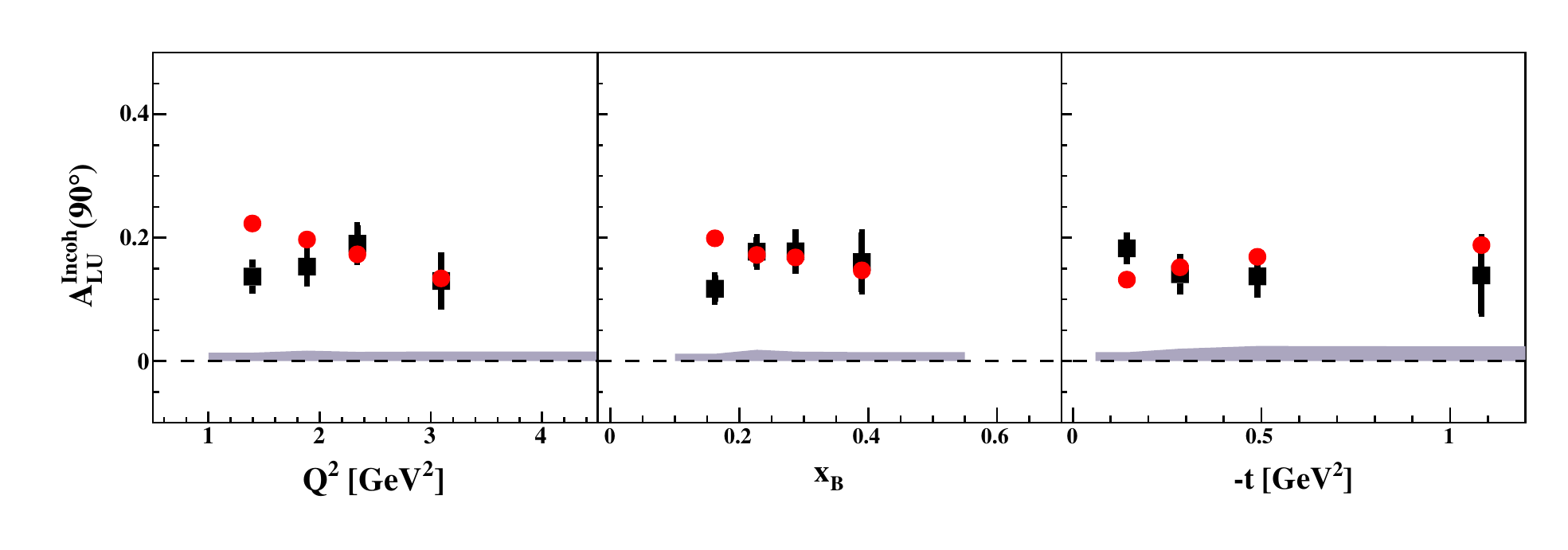}
\caption{
(Color online)
Azimuthal beam-spin asymmetry for the proton in
$^4$He, $A_{LU}^{Incoh}(K)$,
for $\phi = 90^o$: results of this approach (red dots) compared with data
(black squares)
\cite{Hattawy:2017woc}.
From left to right, the quantity is shown in the experimental
$Q^2$, $x_B$ and $t$ bins, respectively. 
Shaded areas represent systematic errors.
}
\label{aluexp}
\end{figure*}

In order to actually
evaluate Eq. \eqref{alu_ratio},
we need an input
for the proton GPD and for the nuclear spectral function. 
Concerning the nuclear part,
only old attempts exist of obtaining a spectral
function of $^4$He
\cite{Morita:1991ka,trento}.
Its realistic evaluation would require
the knowledge, at the same time, of exact solutions
of the Schr\"odinger equation with realistic nucleon-nucleon potentials and three-body forces
for the $^4$He nucleus and for the three-body recoiling system, which can be also unbound with an excitation
energy $\epsilon^*_{A-1}$.
This latter part represents a very complicated
few-body problem, whose solution
is presently unknown. 
A full realistic calculation
of the $^4$He spectral function is planned
and has started but,
in this work, for
$P^{^4He}(\vec p, E)$
use is made of the model 
presented in Ref. \cite{Viviani:2001wu,Rinat:2004ia}.
In particular, 
when the recoiling system is in its ground state,
an exact description is used
in terms of
variational wave functions
for the 4-body 
and 3-body 
systems,
obtained through the hyperspherical 
harmonics method \cite{hh},
within the Av18 NN interaction \cite{Wiringa:1994wb},
including UIX three-body forces \cite{Pudliner:1995wk}.
The cumbersome part of the spectral function, with the recoiling system excited,
is  
based on the Av18+UIX interaction, proposed
in Ref. \cite{Viviani:2001wu,Rinat:2004ia},
an
update of the two-nucleon correlation model of Ref. \cite{CiofidegliAtti:1995qe}.
We note in passing that a realistic calculation of GPDs for $^3$He has been
completed, where the importance of the $E$-dependence
of the spectral function has been established (see Ref. \cite{noi} and references there in).
Clearly our final results will depend on the adopted model for the spectral function.
We stress anyway that, since other available models are less realistic
than the one exploited here, their use would not add interest to the present analysis. 
As already said, we are actively working to obtain an exact spectral function.

For the nucleonic GPD, the model of
Goloskokov and Kroll (GK) 
\cite{Goloskokov:2006hr} has been used,
as we did succesfully in the coherent case
\cite{Fucini:2018gso}.
We remind that
the model is valid in principle at $Q^2$ values larger than those of 
interest here, in particular at $Q^2 \ge 4$ GeV$^2$. Nonetheless
we checked that the GK model can reasonably describe free proton data collected
in similar kinematical ranges \cite{Girod:2007aa}. We therefore adopt it, also because other workable 
global models of GPDs are not easily available. The study of the dependence 
of our results on proton GPDs models is beyond the scope of the present letter
and will be presented elsewhere \cite{tobe}. 

\begin{figure}[t]
\hspace{-.5cm}
\includegraphics[scale=0.45,angle=0]{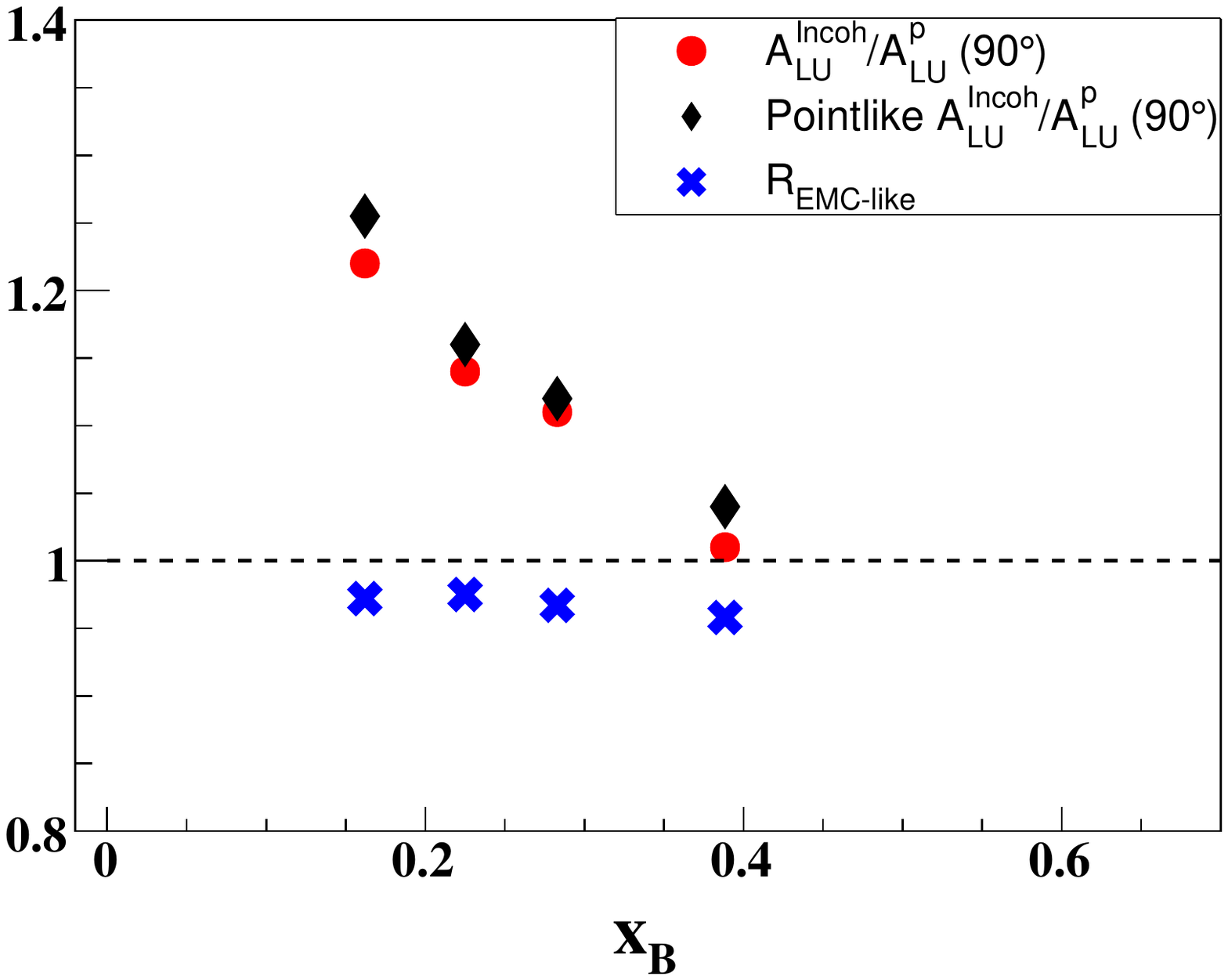} 
\caption{(color online) The ratio 
$A_{LU}^{Incoh} / { A_{LU}^p }$,
Eq. \eqref{aluratio} (red dots),
compared to the result obtained with
pointlike protons (black diamonds)
and to the EMC-like ratio Eq.
\eqref{emclike} (blue crosses).
}
\label{emc}
\end{figure}

With these ingredients at hand, 
Eq. \eqref{alu_ratio} can be evaluated
and the comparison with the recent data \cite{Hattawy:2018liu}
is possible. The BSA 
is a function of the azimuthal angle $\phi$
and of the kinematical variables $Q^2$, $x_B$ and $t$. Due to limited 
statistics, in the experimental analysis
these latter variables have been studied separately 
with a two-dimensional 
data binning. The same procedure has been used in our calculation.
For example, each point at a given $x_B$ has been obtained
using for $t$ and $Q^2$ the corresponding average experimental values. 
In Fig.~\ref{aluexp}, it is seen that,
overall, the calculation reproduces the data
rather well.
In particular, the agreement is not satisfactory only in the region
of low $Q^2$.
Indeed, this is evident only in the experimental
points corresponding to the lowest values
of $Q^2$, $x_B$ and $t$.
One should notice that the average value
of $Q^2$ grows with increasing $x_B$
and $t$, so that a not satisfactory  description
at low $Q^2$ affects also the first $x_B$
and $t$ bins.
A careful analysis of the interplay between
the $t$ and $Q^2$ dependence of the data
is required to establish whether final state interaction (FSI) effects, 
beyond IA,
could be responsible of this disagreement.
In the light of this comparison, we can conclude that
a careful
use of basic conventional ingredients is able to
reproduce the data. 
In order to better understand our results disentangling nuclear modifications possibly related to the EMC effect, as an illustration
we divide our BSA
by the corresponding free proton
quantity, 
as it is given in the literature (see, e.g., Eq. (2) in Ref. 
\cite{Kirchner:2003wt}), 
based on the GK model used in the calculation, and we plot it as a function
of $x_B$.
As it is seen in Fig. 4,
an effect as big as 25 percent is found.
Is that a medium modification of the parton structure?
Actually, such a ratio can be sketched as follows
\begin{equation}
\frac{ A_{LU}^{Incoh} } { A_{LU}^p } 
= \frac{ \mathcal{I}^{^4He} } { \mathcal{I}^{\,p} }
\frac{ T_{BH}^{2\, \, p} } { T_{BH}^{2\, \, ^4He} } \propto \frac
{(nucl.eff.)_{\mathcal{I}}}{(nucl.eff.)_{BH}}\,,
\label{aluratio}
\end{equation}
i.e., it is proportional to the ratio
of the nuclear effects 
on the BH and DVCS interference
to
the nuclear effects on the BH cross section.
If the nuclear dynamics modifies $\mathcal{I}$ and the $T_{BH}^2$
in a different way, the effect
can be big even if the parton structure
of the bound proton does not change appreciably.
Our analysis suggests that this is the case.
This is seen in the other curves presented in
Fig. \ref{emc}. One of them, labelled "pointlike", is obtained considering in the ratio
pointlike protons.
Basically, the big effect is still there.
Besides, in the same figure we show an "EMC-like"
quantity, i.e., a ratio of a nuclear parton observable,
the imaginary part of the CFF, 
to the $same$ observable for the free proton:
\begin{equation}
R_{EMC-like}=\frac{1}{{\cal N}} \frac{\int_{exp}
dE \, d\vec{p} \,P^{^4 He} ( \vec{p}, E) \,\Im m \, \mathcal{ H}(\xi',\Delta^2)}{\Im m \, \mathcal{ H}(\xi,\Delta^2)}  \,,
      \label{emclike}
\end{equation}
where the factor
${\cal N}=\int_{exp} dE \, d\vec{p} P^{^4 He} ( \vec{p}, E)$
accounts for the 
fact that only a part of the
spectral function is selected in a given
experimental bin.
One should notice that this ratio would be one
if nuclear effects $in$ $the$ $parton$ $structure$ were negligible.
As seen in Fig. \ref{emc}, this ratio
is close to one and it resembles the EMC ratio, for $^4$He, at low $x_B$
\cite{Seely:2009gt}.
Since in our analysis the inner structure of the bound proton is entirely contained in the CFF and this produces a mild modification,
the big effect found for the ratio \eqref{aluratio},
shown in Fig. \ref{emc}, has little to do with a modification of the 
parton content. 
Rather, the effect is due to the different dependence on the 4-momentum components, affected by nuclear effects, 
of the interference and BH terms for the bound proton, or to other 
subtle effects.

Our thorough impulse approximation analysis,
based on state-of-the-art models for the
proton and nuclear structure, using a conventional description
in terms of nucleon degrees of freedom, reproduces well
the data on incoherent
DVCS off $^4$He.
This is true especially at high values of $Q^2$;
the disagreement at low $Q^2$ points to possible FSI effects, to be investigated,
or to other quark and gluon effects.
While a benchmark calculation in the kinematics of the
next generation of precise measurements will require
an improved treatment of both the nucleonic and the nuclear
parts, such as a realistic evaluation
of the spectral function and a test of further GPDs models,
with an attempt to estimate the theoretical uncertainty of our predictions,
the straightforward approach proposed here
can be used as a workable framework for the planning of future
measurements. Possible exotic quark and gluon effects in nuclei, not clearly seen within the present 
experimental
accuracy, will be exposed by comparing forthcoming data with our conventional results. 

We warmly thank R. Dupr\'e and M. Hattawy for many
helpful discussions and technical information and explanations on the EG6 experiment,
and E. Voutier for insightful comments on the
manuscript.
This work was supported in part by the STRONG-2020 project of the European Union’s Horizon 2020
research and innovation programme under grant agreement No 824093, and by
the project ``Deeply Virtual Compton Scattering off $^4$He", in the programme FRB of the University
of Perugia.


\begin{thebibliography}{99}

\bibitem{Aubert:1983xm}
  J.~J.~Aubert {\it et al.} [European Muon Collaboration],
  Phys.\ Lett.\  {\bf 123B} 275 (1983).
  
\bibitem{Hen:2013oha}
  O.~Hen, D.~W.~Higinbotham, G.~A.~Miller, E.~Piasetzky and L.~B.~Weinstein,
  Int.\ J.\ Mod.\ Phys.\ E {\bf 22} 1330017
  (2013).
  
\bibitem{Dupre:2015jha}
  R.~Dupré and S.~Scopetta,
  Eur.\ Phys.\ J.\ A {\bf 52} no.6,  159 (2016).

\bibitem{Cloet:2019mql} 
  I.~C.~Cloët {\it et al.},
  J.\ Phys.\ G {\bf 46}, no. 9, 093001 (2019).

\bibitem{gpds} 
  D.~Müller, D.~Robaschik, B.~Geyer, F.-M.~Dittes and J.~Hořejši,
  Fortsch.\ Phys.\  {\bf 42}, 101 (1994);
  X.~D.~Ji,
  Phys.\ Rev.\ Lett.\  {\bf 78}, 610 (1997);
  A.~V.~Radyushkin,
  Phys.\ Lett.\ B {\bf 380}, 417 (1996).
  



\bibitem{Diehl:2003ny} 
  M.~Diehl,
  Phys.\ Rept.\  {\bf 388}, 41 (2003);
  A.~V.~Belitsky and A.~V.~Radyushkin,
  Phys.\ Rept.\  {\bf 418}, 1 (2005).
  
  
  
\bibitem{Berger:2001zb}
  E.~R.~Berger, F.~Cano, M.~Diehl and B.~Pire,
 Phys.\ Rev.\ Lett.\  {\bf 87} 142302
 (2001). 

\bibitem{Polyakov:2018zvc} 
  M.~V.~Polyakov and P.~Schweitzer,
  Int.\ J.\ Mod.\ Phys.\ A {\bf 33}, no. 26, 1830025 (2018)\,.

\bibitem{Burkardt:2000za} 
  M.~Burkardt,
  Phys.\ Rev.\ D {\bf 62}, 071503 (2000)
  Erratum: [Phys.\ Rev.\ D {\bf 66}, 119903 (2002)]\,.





\bibitem{Airapetian:2009cga} 
  A.~Airapetian {\it et al.} [HERMES Collaboration],
  Phys.\ Rev.\ C {\bf 81}, 035202 (2010).

\bibitem{eric}
H. Egiyan, F.-X. Girod, K. Hafidi, S. Liuti, E. Voutier {\it et al.} Jefferson Lab Experiment {\bf E-08-024} (2008).

\bibitem{Hattawy:2017woc} 
  M.~Hattawy {\it et al.} [CLAS Collaboration],
  Phys.\ Rev.\ Lett.\  {\bf 119}, no. 20, 202004 (2017).

\bibitem{Hattawy:2018liu} 
  M.~Hattawy {\it et al.} [CLAS Collaboration],
  Phys.\ Rev.\ Lett.\  {\bf 123}, no. 3, 032502 (2019).


\bibitem{Armstrong:2017wfw} 
  W.~R.~Armstrong {\it et al.},
  arXiv:1708.00888 [nucl-ex].

\bibitem{Guzey:2003jh} 
  V.~Guzey and M.~Strikman,
  Phys.\ Rev.\ C {\bf 68}, 015204 (2003);
  V.~Guzey, A.~W.~Thomas and K.~Tsushima,
  Phys.\ Lett.\ B {\bf 673}, 9 (2009).

\bibitem{Liuti:2005gi} 
  S.~Liuti and S.~K.~Taneja,
  Phys.\ Rev.\ C {\bf 72}, 032201 (2005);
  Phys.\ Rev.\ C {\bf 72}, 034902 (2005).

\bibitem{Fucini:2018gso} 
  S.~Fucini, S.~Scopetta and M.~Viviani,
  Phys.\ Rev.\ C {\bf 98}, no. 1, 015203 (2018).

\bibitem{Slifer:2008re} 
  K.~Slifer {\it et al.} [E94010 Collaboration],
  Phys.\ Rev.\ Lett.\  {\bf 101}, 022303 (2008).


\bibitem{Belitsky:2001ns}
A.~V.~Belitsky, D.~Mueller and A.~Kirchner,
Nucl.\ Phys.\ B {\bf 629}, 323 (2002).

\bibitem{tobe} 
  S.~Fucini, S.~Scopetta and M.~Viviani,
  in preparation.


\bibitem{DelDotto:2016vkh} 
A.~Del Dotto, E.~Pace, G.~Salmè and S.~Scopetta,
Phys.\ Rev.\ C {\bf 95}, no. 1, 014001 (2017).

\bibitem{htwist} 
  A.~V.~Belitsky and D.~Mueller,
  Phys.\ Rev.\ D {\bf 82}, 074010 (2010).

  
\bibitem{Morita:1991ka} 
  H.~Morita and T.~Suzuki,
  Prog.\ Theor.\ Phys.\  {\bf 86}, 671 (1991).

\bibitem{trento}
  V.~D.~Efros, W.~Leidemann and G.~Orlandini,
  Phys.\ Rev.\ C {\bf 58}, 582 (1998).

\bibitem{Viviani:2001wu} 
  M.~Viviani, A.~Kievsky and A.~Rinat,
  Phys.\ Rev.\ C {\bf 67}, 034003 (2003).

\bibitem{Rinat:2004ia} 
  A.~S.~Rinat, M.~F.~Taragin and M.~Viviani,
  Phys.\ Rev.\ C {\bf 72}, 015211 (2005).

\bibitem{hh}
  A.~Kievsky, S.~Rosati, M.~Viviani, L.~E.~Marcucci and L.~Girlanda,
  J.\ Phys.\ G {\bf 35}, 063101 (2008).




\bibitem{Wiringa:1994wb} 
  R.~B.~Wiringa, V.~G.~J.~Stoks and R.~Schiavilla,
  Phys.\ Rev.\ C {\bf 51}, 38 (1995).

\bibitem{Pudliner:1995wk} 
  B.~S.~Pudliner, V.~R.~Pandharipande, J.~Carlson and R.~B.~Wiringa,
  Phys.\ Rev.\ Lett.\  {\bf 74}, 4396 (1995).




\bibitem{CiofidegliAtti:1995qe} 
  C.~Ciofi degli Atti and S.~Simula,
  Phys.\ Rev.\ C {\bf 53}, 1689 (1996).

\bibitem{noi}
  S.~Scopetta,
  Phys.\ Rev.\ C {\bf 79}, 025207 (2009);
  M.~Rinaldi and S.~Scopetta,
  Phys.\ Rev.\ C {\bf 87}, no. 3, 035208 (2013).

\bibitem{Goloskokov:2006hr} 
  S.~V.~Goloskokov and P.~Kroll,
  Eur.\ Phys.\ J.\ C {\bf 50}, 829 (2007).




\bibitem{Girod:2007aa} 
  F.~X.~Girod {\it et al.} [CLAS Collaboration],
  Phys.\ Rev.\ Lett.\  {\bf 100}, 162002 (2008).

\bibitem{Kirchner:2003wt} 
  A.~Kirchner and D.~Mueller,
  Eur.\ Phys.\ J.\ C {\bf 32}, 347 (2003).


\bibitem{Seely:2009gt} 
  J.~Seely {\it et al.},
  Phys.\ Rev.\ Lett.\  {\bf 103}, 202301 (2009).

\end{thebibliography}
\end{document}